# Two-Bus Holomorphic Embedding Method-based Equivalents and Weak-Bus Determination

Shruti Rao, *Student Member, IEEE*, Daniel Tylavsky, *Senior Member, IEEE*, Weili Yi, *Student Member, IEEE*, Vijay Vittal, *Fellow, IEEE*, Di Shi, *Senior Member, IEEE*, Zhiwei Wang, *Member, IEEE*

*Abstract*--A new method of solving the power-flow problem, the holomorphically embedded load-flow method (HELM) is theoretically guaranteed to find the high-voltage solution, if one exists, up to the saddle-node bifurcation point (SNBP), provided sufficient precision is used and the conditions of Stahl's theorem are satisfied. Sigma ($\sigma$) indices, have been proposed as estimators of the distance from the present operating point to the SNBP, and indicators of the weak buses in a system. In this paper, it is shown that the sigma condition proposed in [2] will not produce reliable results and that a modified requirement can be used to produce a tight upper bound on the SNBP. Introduced is an approach to estimate the weak buses in the system using the HEM power series with numerical results compared to traditional modal analysis for a 14-bus system.

## I. Physical Interpretation of Sigma Indices

THE HELM, proposed by Dr. Antonio Trias, involves a specific implementation of the holomorphic embedding method (HEM) applied to the power-flow problem and is theoretically guaranteed to converge to the operable solution, if it exists, for any given power-flow problem [1]. In a later patent, indices called $\sigma$ algebraic approximants were proposed to be used to estimate the distance from the present operating point to the SNBP of a system and to detect the weak buses in the system, buses that directly impact the voltage stability limit [2]. This letter explains theoretically why the procedure is unreliable, provides numerical results to explain the shortcomings, and provides an alternative theoretically-grounded approach.

The idea behind the $\sigma$ indices is to, in essence, develop a two-bus nonlinear equivalent for each bus of the power system, spanning that bus and the slack bus [2], an equivalent that preserves the slack bus voltage and the voltage at the retained bus. The parameters for this proposed reduced equivalent are calculated to be consistent with the simple two-bus system comprised of a slack bus and a PQ bus as shown in Fig. 1, where $Z$ is the line impedance, $S$ is the complex-power injection at the PQ bus, $V_0$ is the slack bus voltage and $V$ is the PQ bus voltage.

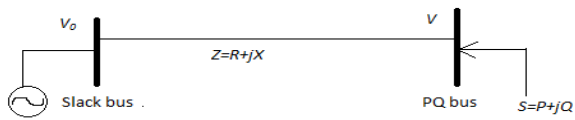

Fig. 1 Two-bus system diagram

The power balance equation for the PQ bus in Fig. 1, given by (1), can be re-arranged to obtain (2),

$$\frac{V - V_0}{Z} = \frac{S^*}{V^*} \qquad (1)$$

$$U = 1 + \frac{\sigma}{U^*}. \qquad (2)$$

where $U=V/V_0$ is the normalized voltage and $\sigma$ is defined as

$$\sigma = \frac{ZS^*}{|V_0|^2}. \qquad (3)$$

The roots of (2), which is a quadratic equation, are:

$$U = 0.5 \pm \sqrt{0.25 + \sigma_R - \sigma_I^2} + j\sigma_I \qquad (4)$$

where $\sigma_I$ and $\sigma_R$ are the imaginary and real parts of $\sigma$, respectively.

If the slack bus voltage is assumed to be controlled at 1.0 pu, the two roots represent the high- and low-voltage solutions for the given two-bus network. The two solutions meet at the point at which the radicand becomes zero, i.e. the SNBP. To ensure the existence of the high-voltage solution in a two-bus system, it is necessary that the radicand in (4) be positive. Thus, the condition to ensure that the two-bus system is short of or at its static voltage collapse point, called the '$\sigma$ condition', is given by:

$$0.25 + \sigma_R - \sigma_I^2 \geq 0. \qquad (5)$$

For a multi-bus system, [2] in essence finds, for all system buses, two-bus equivalents structurally equivalent to Fig. 1 spanning the slack bus and bus of interest, where the PQ bus power injection is the native injection at this bus in the full network. One immediately sees a problem when trying to map a voltage-preserving two-bus *equivalent* (described below) onto the structure of Fig. 1: In a realistic system model, voltages with real parts less than 0.5 are common, but cannot occur for the high voltage solution in a two-bus system (assuming the slack bus voltage angle is 0°) as shown in (4). (In fact, the low voltage solution for the two-bus equivalent constructed below (low voltage solution from (4)) corresponds to the high voltage solution of the full network for cases when the real part of the normalized voltage is below 0.5.) Because [2] does not appear to recognize this incompatibility, nor recognize that taking the low voltage solution would have resolved this incompatibility, their theory leads to erroneous predictions. The approach used by [2] which, in essence, calculates both the equivalent parameters of this two-bus-equivalent analog and the proximity to voltage collapse is explained below along with resultant problematic numerical examples.

The normalized voltages at all buses in a full system model solution can be calculated using HEM as power series of the embedding parameter '$\alpha$', given by $U(\alpha)$. A two-bus equivalent may be constructed, and the $\sigma$ index may then be obtained as a series of $\alpha$, using (6) (which is structurally identical to (2)).

$$U(\alpha) = 1 + \frac{\alpha\sigma(\alpha)}{U^*(\alpha^*)} \qquad (6)$$

The $\sigma$ series can then be evaluated at $\alpha = 1.0$ using Padé approximants [3], to obtain $\sigma$ indices. If a parabola in the complex $\sigma$ plane is plotted as defined by the expression in (5), the distance from the $\sigma$ indices in the plane to the surface of the parabola can be used to qualitatively estimate the distance from the SNBP, and the buses whose $\sigma$ indices are closer to the surface are the weak buses according to [2].

To show it is incorrect to claim that the proximity of the sigma condition to zero is a measure of voltage stability margin [2],

consider the following. The $U$ and $\sigma$ series can be evaluated at $\alpha = 1.0$, where the system loading may be adjusted so that $\alpha = 1.0$ corresponds to any loading condition up to and including the SNBP, and (6) can be split into real and imaginary parts to obtain:

$$\sigma = UU^* - U^*$$
$$\therefore \sigma_R = U_R^2 + U_I^2 - U_R \quad \& \quad \sigma_I = U_I \qquad (7)$$

where $U_R$ and $U_I$ are the real and imaginary parts of the full-model solved bus normalized voltages respectively. The expressions for $\sigma_R$ and $\sigma_I$ from (7) can be substituted into the $\sigma$ condition, (5), to obtain (8).

$$0.25 + \sigma_R - \sigma_I^2 = 0.25 + U_R^2 + U_I^2 - U_R - U_I^2$$
$$= 0.25 + U_R^2 - U_R = (U_R - 0.5)^2 \geq 0 \qquad (8)$$

Thus, it is seen from (8) that the proximity of the $\sigma$ condition to zero, is equivalent to the proximity of the real part of the normalized bus voltage to 0.5. For a simple two-bus (non-equivalent) system, indeed the real part of the normalized voltage is 0.5 at the SNBP and hence the proximity of the $\sigma$ condition to zero is an indicator of the proximity to SNBP. However, for a multi-bus system (and hence its proper two-bus *equivalent*), this is not true and consequently, the magnitude of the $\sigma$ condition is not a measure of closeness to the SNBP, but measures closeness to the point where the alternative root of (4) should be selected. Nor is the radicand of (4) approaching zero a reliable indicator of a weak bus as claimed in [2]; however, the buses whose $\sigma$ conditions first approach zero as the system load increases, have larger phase angles, which cause the real part of the normalized voltage to drop below 0.5.

## II. Numerical Results

The inadequacy of the proposed approach for estimating the SNBP when the system loading is lower than the SNBP loading will be demonstrated numerically using the IEEE 118 bus system. The SNBP of this system occurs when all injections are multiplied by 3.18, i.e., at $\lambda = 3.18$, as obtained from MATPOWER [5], where $\lambda$ is the scaling factor for all complex-power injections in the system. The complex $\sigma$ indices at $\lambda = 1.88$ are plotted for all the buses in the complex $\sigma$ plane and shown in the left plot of Fig. 2. At $\lambda = 1.88$, which corresponds to 60% of the maximum allowable load-scaling factor, many of the $\sigma$ indices are very close to the surface of the parabola, i.e., limit imposed by the $\sigma$ condition. This occurs because some of the normalized bus voltages have real parts very close to 0.5. As the system loading increases further, the real parts of the full-model normalized voltages decrease below 0.5, and hence the numerical value of the expression in (8) increases, causing the $\sigma$ indices to move away from the surface of the parabola. This behavior is shown in the right plot of Fig. 2, where the $\sigma$ indices are plotted at $\lambda = 3.1$, a point much closer to the SNBP.

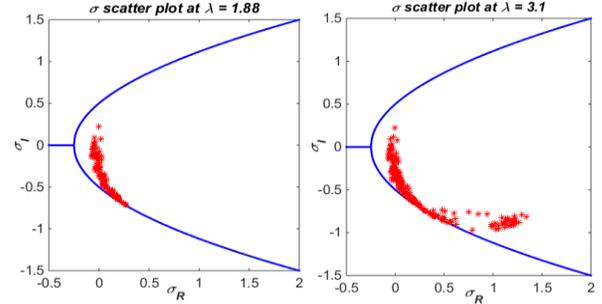

Fig. 2 Plot of $\sigma_I$ vs. $\sigma_R$ at $\lambda = 1.88$ and $\lambda = 3.1$ for the 118-bus system

This behavior is confirmed by plotting the $\sigma$ condition expression against $\lambda$ as shown in Fig. 3 for the first 10 buses (based on native bus numbering) of the 118-bus system. It is seen that some of the buses come very close to violating the $\sigma$ condition as $\lambda$ increases, far before the SNBP is reached, and then start increasing in value again, which agrees with Fig. 2.

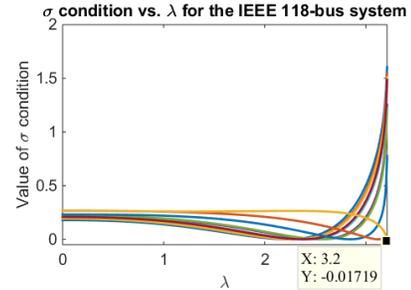

Fig. 3 Plot of $\sigma$ condition vs. $\lambda$ for the 118-bus system

If one searches for the point at which the $\sigma$ condition is *nearly* violated (near zero), one will find that to occur for different buses at loading conditions well below the SNBP, and hence using the proximity of the $\sigma$ condition to zero to estimate closeness to the SNBP would be misleading; however, if one searches for the smallest loading at which the $\sigma$ condition becomes negative (which is theoretically impossible for voltage values short of the SNBP loading) using the Padé approximant of $\sigma(\alpha)$, one will obtain a reasonable estimate of the SNBP for the following reasons (provided a scalable form of the HEM formulation is used [4]).

The Padé approximants of the bus voltages are found to be largely monotonic up to the SNBP, beyond which the voltage function no longer exists, and the evaluation of Padé approximants begins to oscillate wildly due to the poles and zeros (Stahl's compact set) of the approximants. The location of closest real-valued pole/zero when surveyed over all system bus voltages has been found to be a tight upper bound on the true SNBP [4]. Beyond the SNBP, the oscillations in the Padé approximants of the bus voltages (which have no physical interpretation) also appear as oscillations in the $\sigma$ index (oscillations in $\sigma(\alpha)$ vs. $\alpha$). These oscillations inevitably lead to a $\sigma$ condition which becomes negative just beyond the SNBP. Thus the lowest load at which a negative value occurs in the $\sigma$ condition, when surveyed over all buses in the system, is a good indicator of the location of the SNBP. The search method proposed by the authors can be thought of as being analogous (not equivalent) to solving multiple power-flow problems, at increasing load levels, while searching for the point at which the power-balance equations are not satisfied. When the modeled operating condition is even infinitesimally beyond the SNBP, the voltage functions do not exist but the Padé approximants of those voltage functions contain information that can be used to advantage. Beyond the SNBP the high power-balance mismatches are an indication that the system is now modeled to be beyond the SNBP. While it is true that the voltage function does not exist at that point, the unacceptable power-balance mismatches can nevertheless be used as an indirect indication of this non-existence. We have validated this on systems with up to 6000 buses [4].

## III. A Different Approach to Estimate the Weak Buses in a System, Using HEM

The traditional method of estimating the weak buses in a system is modal analysis which involves calculating the eigenvalues of the portion of the reduced Jacobian that retains only the Q-V relationships and determining the buses that have the highest participation factors in the critical modes with smallest eigenvalues [6]. One can use the HEM direction-of-change scaling formulation [4] to estimate the weak buses (without needing to perform the eigenvalue analysis required in [6]) by scaling only the incremental reactive power injection of one bus at a time and calculating the sensitivity of its voltage magnitude at that bus with respect to its incremental reactive power injection, given by (9) (for PQ buses). Note that it is necessary that the incremental reactive power injection be the same for all buses, in order to ensure a fair comparison and in this work it is assumed to be 1 MVAr (1.0 pu on a 1 MVA base). Since the reactive power injection $Q_i(\alpha)$ for PQ buses using direction-of-change scaling is given by $(Q_i+\alpha)$, where $Q_i$ is the native reactive power load, its derivative w.r.t. $\alpha$ is 1.0. The sensitivity of the voltage magnitude with respect to the reactive power injection for the $i^{th}$ bus at the given operating point is obtained by evaluating (9) at $\alpha = 0$ (since $\alpha = 0$ corresponds to the base-case in the direction-of-change scaling formulation).

$$\frac{\partial |V_i(\alpha)|}{\partial Q_i(\alpha)} = \frac{\frac{\partial |V_i(\alpha)|}{\partial \alpha}}{\frac{\partial Q_i(\alpha)}{\partial \alpha}} = \frac{\partial |V_i(\alpha)|}{\partial \alpha}, i \in m \qquad (9)$$

This process is repeated for all buses, with a new direction-of-change for each bus that scales only the incremental reactive power injection at that bus in order to obtain the sensitivities. One does not always expect a perfect one-to-one correspondence between the order of the weakest to the strongest buses from modal analysis and the V-Q sensitivities of (9), since modal analysis method does not provide the order of the buses with the highest to the lowest sensitivity but instead provides the order of buses for which the lowest eigenvalue has the largest *contribution* to its V-Q sensitivity. However, as the smallest eigenvalue decreases (i.e., as the system load increases), the V-Q sensitivities at the buses with higher participation factors for the weakest mode depend to a greater extent on this smallest eigenvalue. Hence near the SNBP, and when the smallest eigenvalue is significantly numerically smaller than the next greatest eigenvalue, one can expect a strong correlation between the buses with high V-Q sensitivities and those with high participation factors in the weakest mode. For the 14-bus system modified to contain only PQ buses, the top five weakest buses (in decreasing order of "weakness") obtained using modal analysis (using bus participation factors for the smallest eigenvalue) and the top five buses with highest V-Q sensitivities (in decreasing order of sensitivity) obtained using the HEM direction-of-change scaling formulation are listed in Table 1 at the base-case and when the system is very close to the SNBP. Note that for the HEM method, the buses with positive reactive power injections (i.e., with local VAr support) are not considered, since they are unlikely to be the weak buses of the system from a steady-state voltage stability perspective. Note that the order of the buses changes with the system operating condition, which is expected since both methods are linearized about the given operating point. Also note that the order of the top five weakest buses obtained using both modal analysis and V-Q sensitivities, (9), obtained using the HEM is identical for both operating conditions. The authors also tested the proposed method at five other operating conditions for this system and observed perfect correspondence in all cases.

TABLE 1 Weak Bus Determination Using Direction-of-Change HEM Scaling Formulation and Modal Analysis

| Loading level | Weakest buses (modal analysis) | Weakest buses (HEM) | Smallest eigenvalue |
|---|---|---|---|
| Base-case | 14, 12, 13, 11, 10 | 14, 12, 13, 11, 10 | 0.31847 |
| All base-case $Q_i \times 3.93$ | 14, 10, 13, 9, 11 | 14, 10, 13, 9, 11 | 0.017966 |

At this point, some remarks about the relative complexity of the HEM versus modal analysis approaches are warranted. When using HEM to determine the bus sensitivity at a given operating point, one needs to calculate only the constant term of the power series for $\partial|V_i(\alpha)|/\partial Q_i(\alpha)$ (since it is evaluated at $\alpha = 0$). To do this, one needs to calculate only two terms in the voltage power series, since at $\alpha = 0$, the derivative $\partial|V_i(\alpha)|/\partial \alpha$ depends only on the $\alpha^1$ term in $V_i(\alpha)$. This is done for all the buses with a different direction-of-change for each bus, however calculations are simple and the computation for different buses is completely parallelizable. Note that the constant term of $V_i(\alpha)$ would be the same for all direction-of-change cases since it depends only on the base-case operating condition, which is the same irrespective of the direction-of-change of the scaling. Also in order to get the second term in the series in each case, only a linear system of equations is solved. Additionally, since only the constant term of $\partial|V_i(\alpha)|/\partial Q_i(\alpha)$ is used, Padé approximants are not needed. By comparison, in modal analysis, one needs to calculate the smallest five to ten eigenvalues and the corresponding eigenvectors for the reduced Jacobian matrix, since the mode associated with the minimum eigenvalue at a given operating condition, may not be the most troublesome mode at all operating conditions. Note that there are many methods to selectively calculate the smallest few eigenvalues of a matrix efficiently [6]. However, the HEPF direction-of-change scaling formulation has the advantage that one can calculate the V-Q sensitivities for any desired direction-of-change scaling (such as determining the sensitivity of the voltage magnitudes of all buses when the injections at a given set of buses changes) and the resultant rational approximant is valid for *all* operating conditions through to the SNBP (provided a sufficient number of series terms and Padé approximants are used). The key computational advantage is that the V-Q sensitivities at *all* operating conditions through to the SNBP are obtained by solving a single power-flow problem.

In summary, a simple modification of $\sigma$ condition (search for negative radicand rather than proximity to zero) can be used to estimate a tight upper bound on the SNBP with reasonable accuracy. This condition is satisfied *only* when the modeled load is such that the $U$ and $\sigma$ series no longer have a physical interpretation. Additionally, a different approach to estimating the weak buses in the system using HEM is proposed which is efficient and correlated well with traditional modal analysis, as demonstrated numerically for the 14-bus system. However, the HEPF direction-of-change scaling formulation has the advantage that one can calculate the *V-Q* sensitivities at *all* operating conditions through to the SNBP, by solving a single power-flow problem.

## IV. Acknowledgment

The authors would like to thank Prof. V. Ajjarapu of Iowa State University for his invaluable feedback provided during the preparation of this manuscript.